\documentclass[twocolumn,trackchanges]{aastex63}
\usepackage{lineno}
\usepackage{soul}
\shorttitle{Helium for HAT-P-18b}
\shortauthors{Paragas et al.}

\graphicspath{{./}{figures/}}

\begin{document}

\title{Metastable Helium Reveals an Extended Atmosphere for the Gas Giant HAT-P-18b}

\correspondingauthor{Kimberly~Paragas}
\email{kparagas@wesleyan.edu}

\author{Kimberly~Paragas}
\affiliation{Astronomy Department, Wesleyan University,
96 Foss Hill,
Middletown, CT 06459, USA}

\author[0000-0003-2527-1475]{Shreyas~Vissapragada}
\affiliation{Division of Geological and Planetary Sciences, California Institute of Technology,
1200 East California Blvd,
Pasadena, CA 91125, USA}

\author{Heather~A.~Knutson}
\affiliation{Division of Geological and Planetary Sciences, California Institute of Technology,
1200 East California Blvd,
Pasadena, CA 91125, USA}

\author[0000-0002-9584-6476]{Antonija~Oklop{\v{c}}i{\'c}}
\affiliation{Anton Pannekoek Institute of Astronomy, University of Amsterdam, Science Park 904, 1098 XH Amsterdam, Netherlands}

\author[0000-0003-1728-8269]{Yayaati~Chachan}
\affiliation{Division of Geological and Planetary Sciences, California Institute of Technology,
1200 East California Blvd,
Pasadena, CA 91125, USA}

\author[0000-0002-0371-1647]{Michael~Greklek-McKeon}
\affiliation{Division of Geological and Planetary Sciences, California Institute of Technology,
1200 East California Blvd,
Pasadena, CA 91125, USA}

\author[0000-0002-8958-0683]{Fei~Dai}
\affiliation{Division of Geological and Planetary Sciences, California Institute of Technology,
1200 East California Blvd,
Pasadena, CA 91125, USA}

\author[0000-0002-1481-4676]{Samaporn~Tinyanont}
\affiliation{Department of Astronomy and Astrophysics, University of California, Santa Cruz, CA 95064, USA}

\author[0000-0002-1871-6264]{Gautam~Vasisht}
\affil{Jet Propulsion Laboratory, California Institute of Technology, 4800 Oak Grove Dr, Pasadena, CA 91109, USA}

\begin{abstract}
The metastable helium line at 1083~nm can be used to probe the extended upper atmospheres of close-in exoplanets and thus provide insight into their atmospheric mass loss, which is likely to be significant in sculpting their population. We used an ultranarrowband filter centered on this line to observe two transits of the low-density gas giant HAT-P-18b, using the 200” Hale Telescope at Palomar Observatory, and report the detection of its extended upper atmosphere. We constrain the excess absorption to be $0.46\pm0.12\%$ in our 0.635~nm bandpass, exceeding the transit depth from the \textit{Transiting Exoplanet Survey Satellite} (\textit{TESS}) by $3.9\sigma$. 
% When we interpret this absorption signal as arising from an outflowing atmosphere, we infer an atmospheric mass loss rate
If we fit this signal with a 1D Parker wind model, we find that it corresponds to an atmospheric mass loss rate between $8.3^{+2.8}_{-1.9} \times 10^{-5}$ $M_\mathrm{J}$/Gyr and $2.63^{+0.46}_{-0.64} \times 10^{-3}$ $M_\mathrm{J}$/Gyr for thermosphere temperatures ranging from 4000~K to 13000~K, respectively. With a \textit{J} magnitude of 10.8, this is the faintest system for which such a measurement has been made to date, demonstrating the effectiveness of this approach for surveying mass loss on a diverse sample of close-in gas giant planets.
\end{abstract}


\section{Introduction} \label{sec:intro}
Close-in exoplanets are exposed to high-energy radiation from their host stars, which can lead to atmospheric mass loss. This atmospheric escape appears to shape the observed short-period exoplanet population \citep[e.g.][]{Lopez13, Owen13, Fulton17}, but there are relatively few published measurements of present-day mass loss rates for close-in planets. Prior to 2018, most studies of atmospheric escape used the hydrogen Lyman-$\alpha$ line at UV wavelengths, the H$\alpha$ line at optical wavelengths, and metal lines at UV and optical wavelengths \citep[e.g.,][]{Vidal-Madjar03, Vidal-Madjar04, Jensen12, Yan18, Cauley19}. In recent years, new theoretical and observational work on the helium (He) 1083 nm line have shown that it can also be used for atmospheric mass loss studies. For planets with a sufficient population of metastable helium atoms in their (potentially escaping) upper atmospheres, the 1083~nm line is optically thick at low pressures (high altitudes), increasing their measured transit depths in this line by a factor of a few \citep{Oklopcic18}. K-type stars are favorable targets for metastable helium observations, as they emit relatively low amounts of mid-UV flux (which depopulates the metastable state), while emitting relatively high levels of EUV flux \citep[which populates the metastable state via ground-state ionization and subsequent recombination;][]{Oklopcic19}. 

%Close-in gas giants are expected to exhibit the strongest He I absorption signals, as they have relatively large planet-star radius ratios and are predicted to have significant atmopsheric outflows. 
Excess He I absorption was first detected in the atmosphere of the sub-Saturn WASP-107b using the \textit{Hubble Space Telescope} \citep[\emph{HST};][]{Spake18}. Since then, this line has been used to detect extended atmospheres in six other planets (5 gas giants and 1 sub-Neptune, with masses ranging from $0.044M_\mathrm{J}$ to $1.116M_\mathrm{J}$) using both space and ground based facilities \citep{Allart18, Mansfield18, Nortmann18, Salz18, AlonsoFloriano19, Ninan20, Palle20}. Understanding how the metastable helium signal scales with stellar EUV flux, planetary gravity, and stellocentric distance on a sample of large, well-characterized planets is a prerequisite for He I observations of smaller planets, where measurements and interpretation are inherently more difficult \citep{Kasper20}. Additionally, detections of outflowing gas giant atmospheres in the He I line provide useful constraints on crucial radiative and collisional processes in theoretical atmospheric mass loss models \citep[e.g.][]{Salz16, Oklopcic18}.
%, which can provide insight on the evolutionary process of the exoplanet population. 

Using the He 1083 nm line, we study the extended atmosphere of HAT-P-18b \citep{Hartman11}, which is a Jupiter-sized ($0.947\pm{0.044}R_\mathrm{J}$), Saturn-mass ($0.196\pm{0.008}M_\mathrm{J}$), $T_\mathrm{eq} = 841\pm15$~K planet \citep{Esposito14} orbiting a K2-type star with $J=10.8$ \citep{Cutri03}. The low density of the planet combined with the spectral type of the host star makes it an excellent target for metastable helium studies. Previous Rossiter-McLaughlin studies of this system have shown it to be one of the only planets around a cool main-sequence star with a retrograde orbit \citep{Esposito14}. Additionally, optical transmission spectroscopy has revealed a Rayleigh scattering slope for this planet, potentially due to a high-altitude haze \citep{Kirk17}. A secondary eclipse of HAT-P-18b has also been detected by the \textit{Spitzer} Space Telescope, and the resulting brightness temperature for the planet suggests efficient day-night circulation and/or a nonzero albedo \citep{Wallack19}. 

In this work, we characterize the atmospheric mass loss of HAT-P-18b for the first time. We observed two transits of HAT-P-18b with the Hale 200” Telescope at Palomar Observatory using an ultranarrowband filter centered on the helium 1083~nm line \citep{Vissapragada20a} with the Wide-field InfraRed Camera \citep[WIRC;][]{Wilson03}. Additionally, we use TESS data from Sectors 25 and 26 to improve the ephemeris for this planet and determine its broadband optical transit depth as a comparison for our helium measurement. In Section~\ref{sec:obs} we describe the WIRC and TESS observations, and in Section~\ref{sec:modeling} we jointly model the light curves from both instruments. We describe our results in Section~\ref{sec:results}, and offer some concluding thoughts in Section~\ref{sec:conclusion}.

\begin{figure*}[ht!]
    \centering
    \gridline{\fig{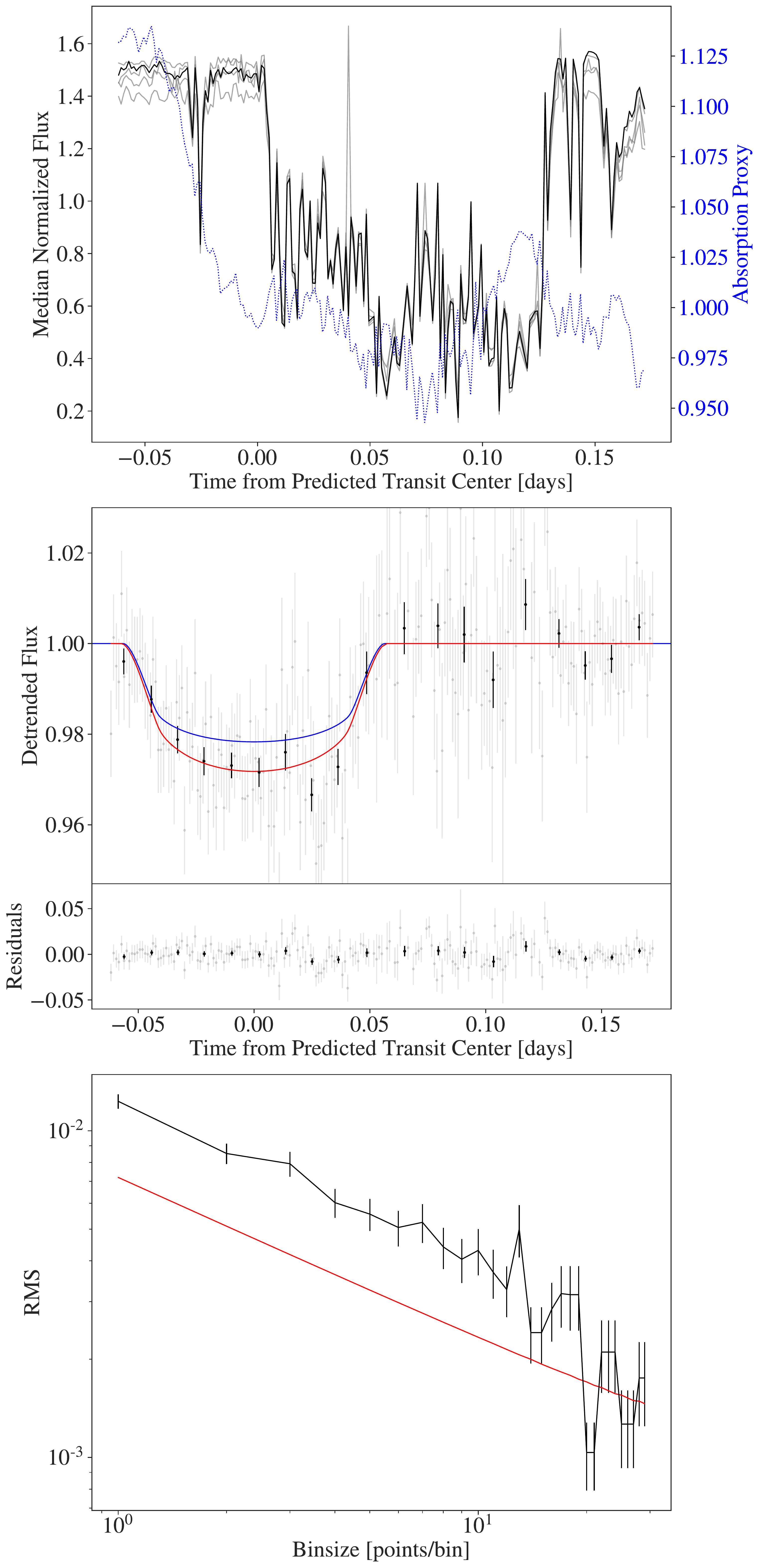}{0.49554\textwidth}{(a)}
          \fig{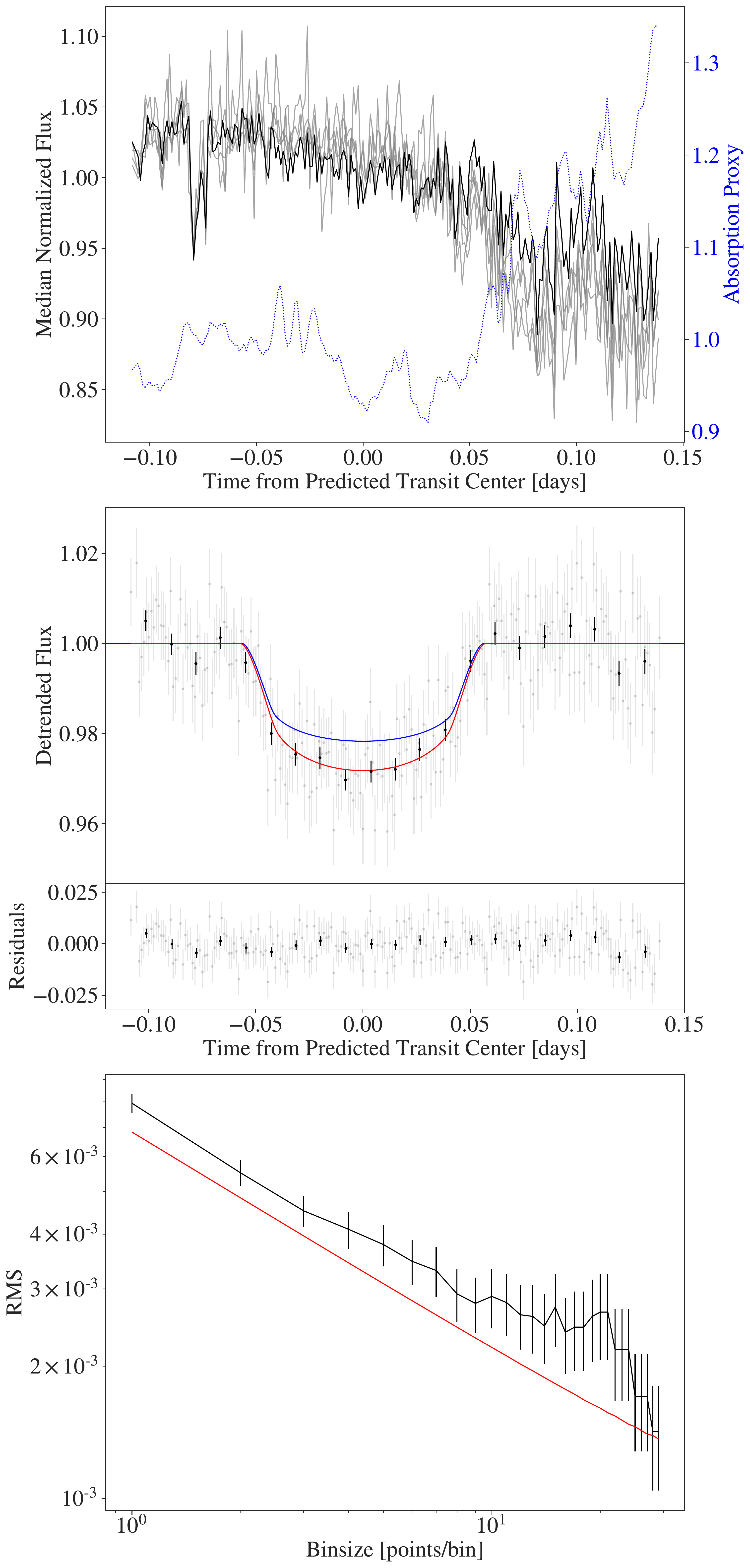}{0.49\textwidth}{(b)}}
          
    \caption{(a) Results for WIRC night 1 and night 2 shown in (a) and (b), respectively. The top row shows the median-normalized raw light curves of the target (in black) and its comparison stars (in grey) as well as the water absorption proxy described in Section \ref{WIRCobs} (dotted blue line) throughout each respective night. The middle row shows the helium light curve with unbinned data in grey and binned data to a 15 minute cadence in black, with the best-fit joint helium model in red and the TESS model in blue, and the residuals of each fit. The bottom row shows the Allan deviation plot for each dataset.}
    \label{plots}
\end{figure*}

\begin{figure*}[ht!]
    \centering
        \includegraphics[width=180mm]{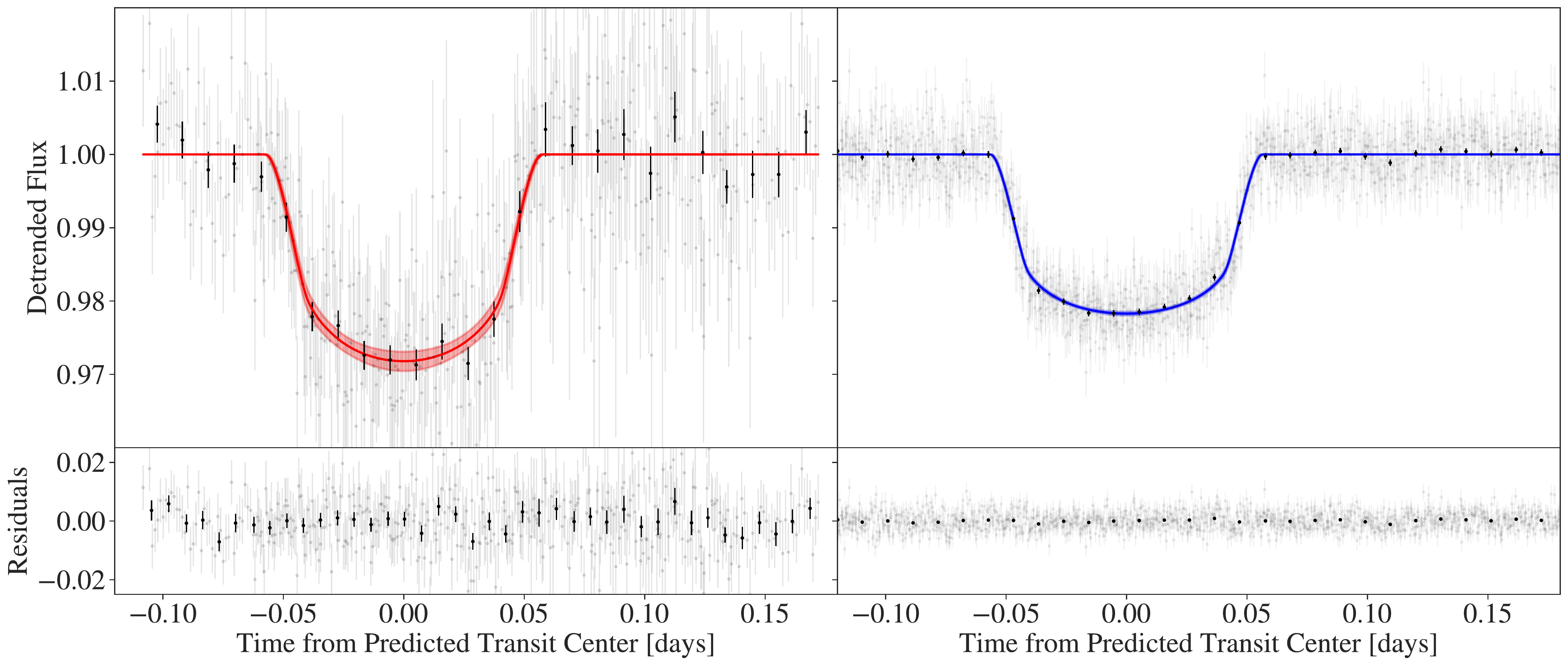}
    \caption{Combined transit light curves and residuals for WIRC (left) and TESS (right), with unbinned data in grey and binned data to a 15 minute cadence in black. The best fit models for WIRC (red) and TESS (blue) are overplotted with the 1$\sigma$ confidence interval denoted by the shaded region.}
    \label{joint}
\end{figure*}

\section{Observations}
\label{sec:obs}
\subsection{WIRC Observations} \label{WIRCobs}

We observed transits of HAT-P-18b on UT June 6 2020 (hereafter night 1) and July 8 2020 (night 2). We used an ultranarrowband filter centered on the helium 1083.3 nm line with a FWHM of 0.635~nm \citep{Vissapragada20a}. For these observations we typically use a custom beam-shaping diffuser, which produces a stable 3$\arcsec$ diameter top hat point spread function \citep{Stefansson17}, to mitigate time-correlated systematics \citep{Vissapragada20b}. However, on night 1 the weather conditions were poor and we elected to defocus the telescope to 1$\farcs$2 rather than using our diffuser, as the precision of the photometry would have been compromised by the increased sky background introduced into the diffused point spread function of the star. On night 2, we utilized the diffuser as normal. Before each observation we observed a helium arc lamp installed at the Hale 200" to determine our region of maximal sensitivity to the 1083~nm line, and we placed our target on this region during data collection. Due to the change in the filter's center wavelength across the detector, telluric OH emission lines form bright radial arcs on the detector; to calibrate this we constructed a background template for the OH lines using a four-point dither on each night. For our night 1 observations, we took 90 second exposures from UT 05:03:30 to UT 10:40:11, beginning at airmass 1.237 and ending at airmass 1.207. For our night 2 observations, we took 90 second exposures from UT 05:06:57 to UT 11:01:56, beginning at airmass 1.011 and ending at airmass 2.301. Both nights reached a minimum airmass of 1.000. 

To calibrate the images, we dark-subtracted and flat-fielded all the science data, correcting for bad pixels and any residual detector striping. This process is described in \citet{Vissapragada20a}. To correct for the bright arcs on the detector caused by telluric OH, we median-scaled the sigma-clipped science data to the dithered background frame in 10-pixel radial steps from the filter zero point (where rays have passed through the filter at normal incidence) at the top of the detector. This process eliminates most of the telluric background, leaving a small residual background that we correct for locally in our aperture photometry process. 

We use these background scaling factors to correct for time-varying telluric water absorption as well. There are two water lines that overlap with the bandpass of our filter at 1083.57~nm and 1083.66~nm. At the effective resolving power of our filter, these two lines appear as a singular absorption line. Previously in \citet{Vissapragada20a}, we could assume that the strength of the absorption in this water line was effectively constant over the night,
%corrected only the OH emission lines and not the nearby water absorption feature,  
as the observing conditions were good and the resulting light curves did not exhibit time-correlated variations corresponding to a rapidly-varying water vapor column. However, since the first of our nights had relatively poor weather conditions with sporadic cloud coverage and noticeable seeing and transparency variations, we sought to track variations in this water absorption feature using the OH sky emission lines. The Lorentzian wing of the $Q_1(3/2)$ OH emission line at 1083.4~nm overlaps with the telluric water feature \citep{Allart18, Salz18}. This means that OH emission originating higher up in the atmosphere \citep[{$>$80~km};][]{Bernath09} can be absorbed by H$_{2}$O while passing through the lower atmosphere. We can therefore track the telluric water variation over the night by dividing the time-varying flux in the water-contaminated OH emission line (as measured by our scaling factors) by that of the uncontaminated  $R_1(3/2)$ and $R_2(1/2)$ OH lines at 1075~nm and 1078~nm.
%When we derive scaling factors to the sky background to remove the OH features, 
If the water variation is significant enough to impact our photometry, we can utilize this absorption proxy as a decorrelation parameter in our transit fits.

We performed aperture photometry on the target star and six comparison stars (the same ones on each night) using the package \texttt{photutils} \citep{Bradley16}. We tested different aperture sizes in one pixel steps from 3 to 13 pixels in radius. We removed $3\sigma$ outliers from the data using a moving median filter. This process is detailed in \citet{Vissapragada20a}. Our optimal aperture sizes for our night 1 and night 2 observations were 7 pixels (1\farcs75) and 11 pixels (2\farcs75) in radius, respectively, with the difference arising from our use of a slight defocus on the first night and a diffuser on the second night. The raw light curves of the target and the comparison stars are shown in Figure~\ref{plots} for both nights. 

\subsection{TESS Observations}
\label{sec:TESSobs}
We used the 2-minute cadence TESS observations of HAT-P-18b obtained during Sectors 25 and 26. TESS observed the target for 51.5 days starting on May 14 2020 and June 9 2020 for Sectors 25 and 26, respectively, covering 8 transits in total. We downloaded the Pre-search Data Conditioning Simple Aperture Photometry (PDCSAP) light curve from the Mikulski Archive for Space Telescopes (MAST) using the \texttt{lightkurve} package \citep{Lightkurve18}. With the transits masked, we removed low-frequency variability from the data using the Savitzky-Golay filter from \texttt{scipy} \citep{Jones01} and rejected $5\sigma$ outliers using a moving median filter. However, we noticed that even after the filter was applied, there were still strong uncorrected systematics that biased the transit depths of the first two transits, so we omitted them from our combined fit. Although these transits may be recoverable with different detrending methods, our constraint on the TESS transit depth from the remaining six transits were sufficiently precise for comparison to the WIRC light curves (i.e., the uncertainty in the comparison is dominated by the uncertainty on the WIRC transit depth).
%has a nearly equivalent precision and is thus sufficient for our purpose here.
% Before we proceeded, we fit the remaining 7 transits with \texttt{exoplanet} \citep{ForemanMackey20} and allowed each transit depth to vary while holding all other light curve parameters constant. We fit each model with \texttt{PyMC3} \citep{Salvatier16} using the No U-Turn Sampler (NUTS) to sample the posterior distributions. From this fit, we noticed that the first transit of the seven had a transit depth that differed from the other 6 transits by greater than 1$\sigma$. This discrepancy is explained by the systematic trend along the light curve, where we see the largest variability at this first transit. This steep trend is not being effectively corrected by the SavGol filter, thus biasing the transit depth value. 

\section{Light Curve Modeling}
\label{sec:modeling}
We simultaneously fit both nights of WIRC data along with the corrected TESS photometry using \texttt{exoplanet} \citep{ForemanMackey20}. For each WIRC light curve, we fit an instrumental noise model consisting of a linear baseline along with a linear combination of comparison star light curves, with the weights of the comparison stars left as free parameters in the fit. This is an update from our previous modeling methodology \citep{Vissapragada20a}, where we used an ordinary least-squares method to quickly determine comparison star weights at each likelihood evaluation, and it is enabled by the rapid No U-Turn Sampler (NUTS) sampler that \texttt{exoplanet} makes available for high-dimensional light-curve fitting. We also tried including two additional parameters in our instrumental noise model for each night: the water absorption proxy (as described in Section~\ref{WIRCobs}), and the distance from the median centroid. We find that two comparison stars in each WIRC night have posterior probability distributions for their weights that overlap with zero, and we therefore remove them from the fit, lowering the Bayesian Information Criterion (BIC) value by 35 and 28 for night 1 and night 2, respectively. Although these stars are not the same for each dataset, the two nights had different observing strategies and weather conditions. We find that we obtain optimal fits when we only include the telluric water proxy in our fits to the night 1 data, lowering the BIC value by 17; this is not surprising, as this night had relatively poor and variable weather conditions. We opted to keep the distance from the median centroid as a decorrelation parameter for both nights, as their removal from the fits resulted in a $\Delta$BIC $<$ 10. Our final systematics model contained 15 parameters: two parameters for each of the linear baselines, four comparison stars for each dataset, the distance from median centroid for each dataset, and the absorption proxy for the first night.

We fit a transit model simultaneously with the systematics. We have three fit parameters that are common to all datasets: the predicted mid-transit time $T_{0}$, the period $P$, and the impact parameter $b$. Initially, we allowed each night of WIRC data to have its own transit depth in the joint fit. The two transit depths ($2.11^{+0.25}_{-0.23}\%$ for night 1 and $2.35\pm{0.14}\%$ for night 2)  were within $1\sigma$ of each other, indicating that the magnitude of helium absorption appears consistent between these two epochs. We therefore fit a single transit depth for both helium light curves. We fit for the limb darkening coefficients $[u_{1},u_{2}]_\mathrm{(He)}$ and $[u_{1},u_{2}]_\mathrm{(TESS)}$ and transit depths $({R_{p}}/{R_{\star}})^{2}_\mathrm{(He)}$ and $({R_{p}}/{R_{\star}})^{2}_\mathrm{(TESS)}$ in each bandpass, both sampled uniformly (see Table~\ref{table1}). For each WIRC light curve, we fit for a jitter parameter describing the excess noise in addition to the photon noise log($\sigma_{extra}$). For TESS, we noticed that the error bars that came with the PDCSAP fluxes were not an accurate representation of the photon noise and in fact overestimated the observed scatter in the data.  We therefore include a scaling factor $k$ for the TESS error bars.

%We account for uncertainties in $M_\star$ and $R_\star$ by allowing these two parameters to vary in our fits with priors from \citet{Esposito14}.  Our data do not further constrain these parameters, so we omit them from our final results and corner plots.

\begin{figure*}[htb!]
    \centering
        \includegraphics[width=180mm]{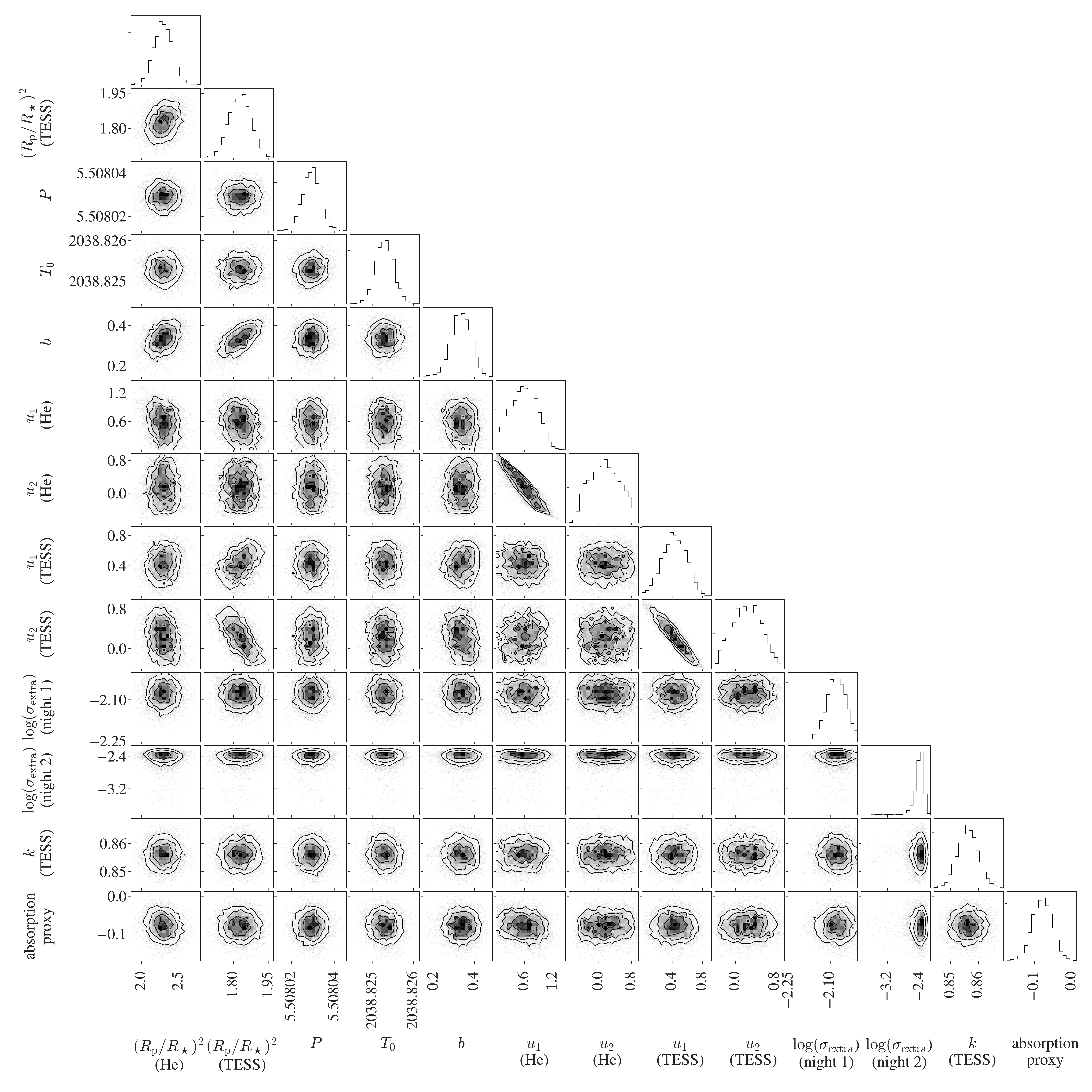}
    \caption{Corner plot displaying the posterior probability distributions for the joint model for HAT-P-18b. Note transit depth ($R_p$/$R_*$)$^2$ values are in \%, period $P$ is in days, and predicted mid-transit time $T_0$ is in BTJD$_\mathrm{TDB}$. We omit all of the decorrelation parameters except for the absorption proxy.}
    \label{corner plot}
\end{figure*}

We use the NUTS in \texttt{PyMC3} \citep{Salvatier16} to sample the posterior distributions for our model parameters. We ran four chains, tuning each for 1500 steps (the ``burn-in'' period) and then taking 1000 draws in each chain, achieving good convergence with a Gelman-Rubin \citep{Gelman92} statistic of $< 1.006$ for all parameters. 
%(While this is far fewer draws than would be required in a typical MCMC, gradient information allows for very rapid convergence in NUTS even for posterior distributions with strong correlations, as evidenced by our effective sample sizes in excess of 1000 for all parameters).
The priors and posteriors for the physical parameters in our model are given in Table~\ref{table1} for the joint fit and the detrended light curve, residuals, and Allan deviation plot for each night of WIRC data are displayed in Figure~\ref{plots}. The final combined helium and TESS light curves are displayed in Figure~\ref{joint}, and the posterior distributions for the model parameters are visualized in Figure~\ref{corner plot}.

\begin{deluxetable*}{cccc}
    \tablecolumns{4} 
    \tablewidth{300pt} 
    \tablecaption{Priors and posteriors for joint fit to Palomar/WIRC and TESS data}
    \tablehead{ \colhead{Parameter} & \colhead{{Prior}} & \colhead{Posterior} & \colhead{Units}}
    \startdata
        $(R_\mathrm{p}/R_\star)^2$ (He) & $\mathcal{U}(1, 25)$ & $2.29\pm{0.12}$ & \% \\
        $(R_\mathrm{p}/R_\star)^2$ (TESS) & $\mathcal{U}(1, 25)$ & $1.832^{+0.045}_{-0.048}$ & \% \\
        $P$ & $\mathcal{N}(5.5080291, 0.0000042)$ & $5.508029\pm{0.0000042}$ & days  \\
        $T_{0}$ & $\mathcal{U}(2038.5, 2039.0)$ & $2038.82530\pm{0.00023}$ & BTJD$_\mathrm{TDB}$ \\
        $b$ & $\mathcal{N}(0.352, 0.057)$ & $0.338^{+0.047}_{-0.051}$ & -- \\
        $u_1$ (He) & \citet{Kipping13} & $0.58^{+0.29}_{-0.30}$ & -- \\
        $u_2$ (He) & \citet{Kipping13} & $0.14\pm{0.39}$ & --\\
        $u_1$ (TESS) & \citet{Kipping13} & $0.45\pm{0.16}$ & -- \\
        $u_2$ (TESS) & \citet{Kipping13} & $0.20^{+0.30}_{-0.31}$ & --  \\
        $log(\sigma_{extra})$ (night 1) & $\mathcal{U}(-4,-2)$ & $-2.078^{+0.036}_{-0.038}$ & -- \\
        $log(\sigma_{extra})$ (night 2) & $\mathcal{U}(-4,-2)$ & $-2.40^{+0.08}_{-0.12}$ & -- \\
        $k$ (TESS) & $\mathcal{U}(0.5, 1.5)$ & $0.8563^{+0.0037}_{-0.0036}$ & -- \\
        absorption proxy (night 1) & $\mathcal{N}(0.0, 0.1)$ & $-0.078^{+0.026}_{-0.028}$ & -- \\
    \enddata
    \tablecomments{BTJD$_\mathrm{TDB}$ = BJD - 2457000. Note that we omitted the stellar parameters and all of the detrending weights except for the absorption proxy for night 1.}
     \label{table1}
\end{deluxetable*}

\begin{figure}[ht!]
    \centering
        \includegraphics[width=85mm]{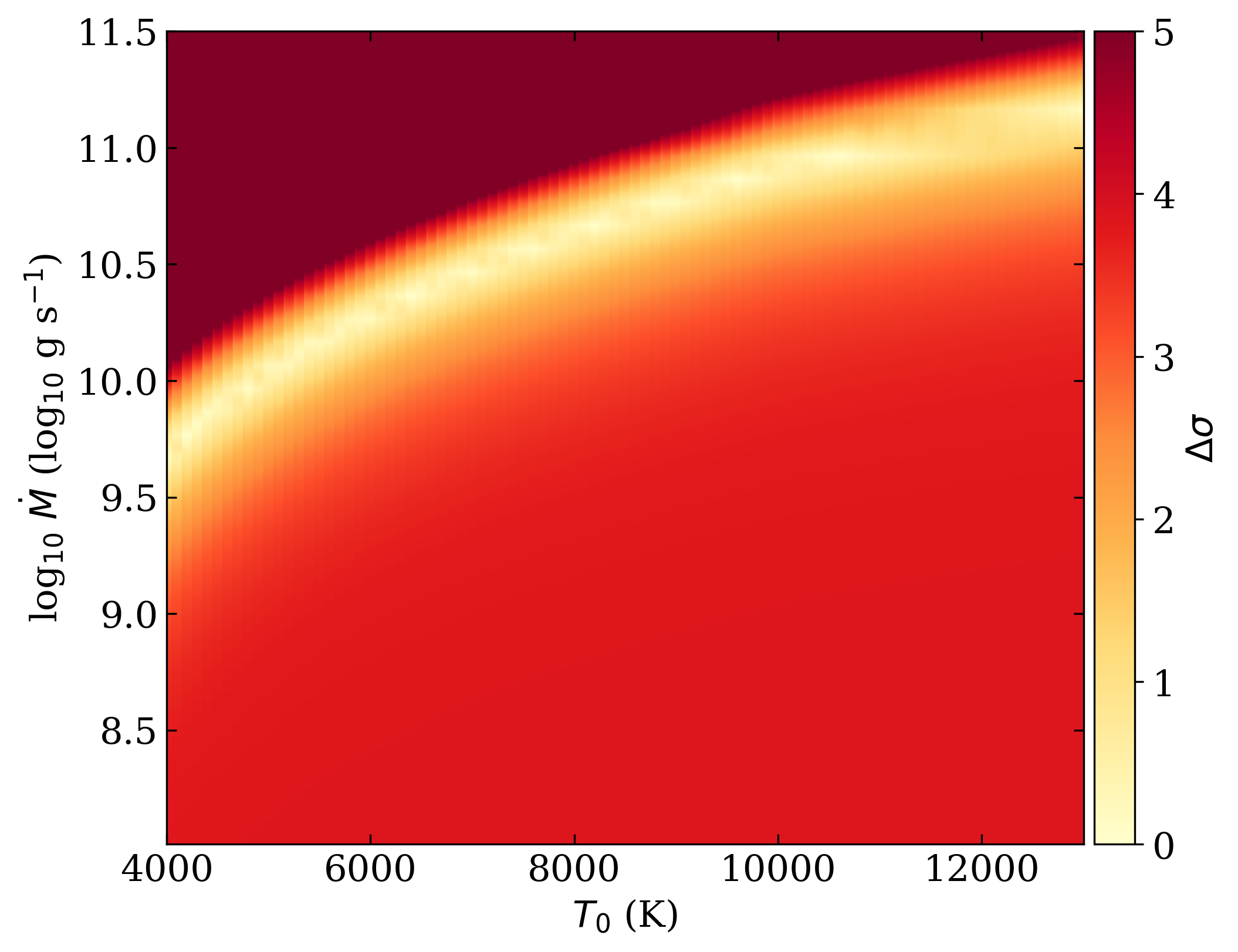}
    \caption{Atmospheric mass loss model for HAT-P-18b. Each point is a different mass loss model corresponding to specific $T_0$ and $\dot{M}$ values, and the shading indicates the compatibility between the model and our observed excess absorption (with the lighter regions indicating the most concordant models).}
    \label{banana plot}
\end{figure}

\section{Results \& Discussion} \label{sec:results}
We measure a transit depth of $2.29^{+0.12}_{-0.13}\%$ in the helium line. The corner plot for our fit parameters is shown in Figure~\ref{corner plot}. Our measurement can be compared to the TESS transit depth measurement of $1.830^{+0.049}_{-0.052}\%$. Our measured transit depth in the helium bandpass exceeds that in the TESS bandpass by $0.46\pm{0.12}\%$ ($3.9\sigma$). The TESS bandpass is between 600~nm to 1000~nm, making it reasonable to use as a comparison for our measurement. Previous studies on our target have shown variation within this range to be limited to variations on the order of the scale height \citep{Kirk17}. The difference between the two transit depths exceeds, by an order of magnitude, the expected change in transit depth for a one (lower-atmospheric) scale-height change in planet radius (0.03\%) for this target. Thus, the observed excess absorption cannot be explained by broad absorption features -- for instance, by water -- in the lower atmosphere. The helium line is near an opacity minimum of water anyways so this explanation is disfavored a priori. We conclude that the observed absorption indeed arises from metastable helium in HAT-P-18b's extended atmosphere.
%This value is most similar to the measured excess absorption of $0.498\pm{0.045}$\% for WASP-69b in \citet{Vissapragada20}, 
%Our target is nearly 3 magnitudes fainter than WASP-69b, which has a measured excess absorption of $0.498\pm{0.045}$\% \citep{Vissapragada20b}, estimated using the same approach as this work, therefore it is expected that our measurement has a correspondingly greater uncertainty.
%HAT-P-18b is the faintest target with detected excess helium absorption to date.

We use the model described in \citet{Oklopcic18} to convert our measured excess absorption into a joint constraint on HAT-P-18b's mass loss rate $\dot{M}$ and upper atmospheric temperature $T_{0}$. This model calculates the velocity and density profiles of a 90\%/10\% H/He 1D Parker wind  as a function of $\dot{M}$ and $T_{0}$, and then calculates the level populations for helium given a UV stellar spectrum. We use the MUSCLES UV spectrum of $\epsilon$ Eridani \citep{France16, Loyd16, Youngblood16}, which is another K2 type star, as a stand-in for the unknown UV spectrum of HAT-P-18. Accounting for the stellar radius and semi-major axis of HAT-P-18b, the EUV irradiance of the planet was 8 W/m$^2$ integrated between 5.5 \AA\ and 911 \AA. The results are shown in Figure~\ref{banana plot}. HAT-P-18b's mass loss rate is likely between $8.3^{+2.8}_{-1.9} \times 10^{-5}$ and $2.63^{+0.46}_{-0.64} \times 10^{-3}$ $M_{J}$/Gyr for thermosphere temperatures between 4000 and 13000~K, respectively. Using the EUV irradiance above along with an efficiency parameter $\varepsilon = 0.1$, we can also calculate an energy-limited mass loss rate for HAT-P-18b \citep[e.g.][]{MurrayClay09}:
\begin{equation}
    \dot{M} = \frac{\varepsilon\pi R_\mathrm{p}^3F_\mathrm{XUV}}{GM_\mathrm{p}} \approx 4\times10^{10}~\mathrm{g/s}.
\end{equation}
This estimate agrees well with the inferred mass loss rates in Figure~\ref{banana plot} (with uncertainties of a factor of few accommodated by similar uncertainties on the efficiency parameter), suggesting that our observationally-derived constraints are energetically feasible.

Because $\epsilon$ Eridani is a relatively young, active star, with $\log(R_\mathrm{HK}) = -4.51$ compared to $\log(R_\mathrm{HK}) = -4.80$ for HAT-P-18, we repeated the modeling using the MUSCLES spectrum of HD 40307, a fairly inactive ($\log(R_\mathrm{HK}) = -4.99$) K2.5V star. By trying proxy stars with activity levels on either side of HAT-P-18b's activity, we can get a sense for the uncertainty on the result  based on our choice of EUV proxy. With HD 40307 as a proxy, we found a best-fit $\log(\dot{M}) = 9.60^{+0.11}_{-0.12}$ at 4000~K, and $11.20^{+0.12}_{-0.11}$ at 13000~K, nearly identical to our findings for $\epsilon$ Eri. This is because the mid-UV to EUV flux ratios between the two stars are quite similar \citep{Oklopcic19}. Although $\epsilon$ Eri is a much stronger X-ray emitter, the cross section to X-ray photoionization of helium is very small compared to the cross section in the EUV near the 504~\AA\ threshold, so the contribution to the flux-averaged photoionization cross section is negligible \citep{Oklopcic18}.

We note that there is a strong degeneracy between the mass loss rate and thermosphere temperature due to the complex dependence of the outflow velocity and density on the temperature and the mass loss rate. This degeneracy could be partially resolved with a precise line shape measurement, but we do not resolve the line shape in these observations. Due to the faintness of HAT-P-18, spectrographs on all but the largest telescopes may have difficulty resolving the line shape precisely enough to break the degeneracy. Additionally, our helium light curve is symmetric across our best-fit mid-transit time. However we cannot exclude the possibility of an extended egress, as our combined light curve lacks the precision required to significantly detect a trailing helium tail for such a faint target.

\section{Conclusions} \label{sec:conclusion}
In this work, we use an ultra-narrowband helium filter centered on the 1083~nm line to observe two transits of HAT-P-18b. We detect $0.46\pm{0.12} \%$ excess helium absorption in the planet's upper atmosphere. This detection corresponds to an atmospheric mass loss rate between $8.3^{+2.8}_{-1.9} \times 10^{-5}$ and $2.63^{+0.46}_{-0.64} \times 10^{-3}$ $M_{J}$/Gyr, which means HAT-P-18b is losing less than 2\% of its mass per Gyr. This is typical for close-in gas giants, with other helium outflow detections having mass loss rates less than 5\% per Gyr \citep{Allart18, Mansfield18, Spake18, AlonsoFloriano19}.

Of the handful of planets with detected helium outflows, WASP-107b is the most comparable to HAT-P-18b with a similar radius of 0.94 $R_\mathrm{J}$, mass of 0.12 $M_\mathrm{J}$, separation of 0.55~au, and equilibrium temperature of $770$~K \citep{Anderson17}. If we assume that HAT-P-18b has a He line shape similar to that of WASP-107b, we can invert our excess helium transit depth to obtain an estimate of the underlying predicted line depth of $4.5 \pm{1.3}\%$. This is noticeably smaller than the $7.26 \pm 0.24\%$ depth measured by CARMENES and Keck/NIRSPEC for WASP-107b \citep{Allart19, Kirk20}. The difference may be due to the smaller gravitational potential of WASP-107b \citep[the mass of WASP-107b has recently been suggested to be even lower by ][]{Piaulet20}, or differences in the EUV spectra of the two stars (WASP-107 is a K6 star while HAT-P-18 is a K2). Detailed comparative modeling of these two planets may make clear the primary control on the metastable helium signal.

%WASP-107b has an estimated mass loss rate between $2 \times 10^{-4}$ and $5 \times 10^{-3}$ $M_{J}$/Gyr \citep{Spake18}; this value is very similar to HAT-P-18b's. These two planets are losing at most 5\% of their total mass per Gyr, indicating that mass loss is unlikely to play a significant role in their long-term evolution. 

This is the faintest system ($J=10.8$) with detected helium absorption thus far, establishing the effectiveness of our technique for observing such targets with a mid-sized telescope. For reference, the next faintest system with detected excess helium absorption is WASP-107b with a $J$ magnitude of 9.4. Of the 11 planets identified in \citet{Kirk20} as promising targets for observations of helium outflows, many are challenging targets with $J > 10.5$. Our photometric technique allows us to begin surveying planets around such faint stars, expanding the sample of planets with measured metastable helium absorption. Further population-level studies of extended atmospheres in the He 1083~line will greatly improve our ability to calibrate the mass loss models used to elucidate the long-term evolution of the close-in exoplanet population.

\acknowledgements
We are thankful to the Palomar staff, especially Kajse Peffer, Paul Nied, Joel Pearman, Carolyn Heffner, and Kevin Rykoski for their support. KP acknowledges support from the Summer Undergraduate Research Fellowship (SURF) at the California Institute of Technology and the NASA CT Space Grant. SV is supported by an NSF Graduate Research Fellowship and the Paul \& Daisy Soros Fellowship for New Americans. HAK acknowledges support from NSF CAREER grant 1555095.

\end{document}